\documentclass{appolb}
\usepackage{epsfig,amsmath,amssymb}
\usepackage{calc,color}
\pdfoutput=1

\newcommand{\ord}{\mathcal{O}}
\def\Mkk{M_{\rm KK}}
\newlength{\figureoverhang}
\begin{document}
\title{Randall-Sundrum Models and Precision Observables%
\thanks{Presented at the FLAVIAnet workshop in Kazimierz: \emph{Low energy
constraints
on extensions
of the Standard Model} }%
}
\author{Martin Bauer
\address{Johannes-Gutenberg Universit\"at Mainz}
%\and
%the Name(s) of other Author(s)
%\address{and their affiliation}
}
\maketitle
\begin{abstract}
I present a review of phenomenological
implications of the Randall-Sundrum (RS) model with bulk fermions and
brane-localized Higgs boson. Modifications to the $W$-boson mass, corrections
to the Peskin-Takeuchi parameters and to the $Z b\bar{b}$ couplings will be
discussed. From these observables severe bounds on the mass scale of
Kaluza-Klein (KK) modes arise. Constraints from all three observables are very
sensitive to the exact value of the Higgs boson mass and the bounds can be
significantly lowered by allowing for a heavy Higgs boson ($m_h\sim 1 $TeV).
Consequences thereof, as well as other approaches like ``little RS'' models and
models with custodial symmetry will also be briefly discussed.
\end{abstract}
  
\section{Introduction}
   Extra dimensional models with a warped background were proposed ten
years ago by Randall and Sundrum (RS) \cite{Randall:1999ee} in order to solve
the gauge hierarchy
problem. In these models the fifth dimension is an $S_1/Z_2$ orbifold, which is
warped due to a non-factorizable metric
\begin{equation}
ds^2 = e^{-2\sigma(\phi)} \eta_{\mu\nu} dx^{\mu}dx^{\nu} - r^2 d\phi^2, \qquad 
\sigma(\phi) = kr|\phi| ,
\end{equation} 
where $k$ denotes the curvature and $r$ the radius of the extra dimension.
The extra dimension is bound by two branes, the ultra-violet brane (UV) at
$\phi = 0$ and the infra-red (IR) brane at $\phi = \pi$.
Due to the warp factor $e^{-2\sigma(\phi)}$, energy scales in this model depend
on the position along the fifth dimension. This allows to address the
gauge hierachy
problem if the Higgs field is located at the IR brane.
Without additional constraints, all other SM fields are allowed to probe the
fifth dimension. In this setup\footnote{I will refer to this model as the
\emph{minimal model} in contrast to extensions with custodial protection.} the
localization of quark fields along the extra dimension provides an attractive
explanation of the flavor puzzle. Section 2 reviews the theoretical
framework of this model. 
 \\
    Many qualitative and quantitative studies have been performed in
the last years and allow for a detailed understanding of the limits on the KK
mass scale.
% the last years analysis of
%warped extra dimensional models have left
%the stage of qualitative implications and reached a phase where quantitative
%computations allow for a detailed evaluation of its limits and prospects.
In Sections 2 and 3 the results of these computations will be summarized with
particular emphasis on electroweak precision observables. Corrections to the
$S, T,$ and $U$ parameter, to the $Zb\bar{b}$ vertex, and to
the mass of the $W$ boson, as well as possible constraints from these
observables will be examined. Consistency with the bounds coming from the
current experimental status of these observables can be achieved within the
minimal model, but also models with an extended gauge group received increasing
attention in the last years, as they provide an elegant solution to the tension
coming especially from the constraints from $Zb\bar{b}$ and $T$. I will
discuss pros and
cons of the different solutions to round off the review. The results of
this proceedings are based on \cite{Casagrande:2008hr} and an extended study of
flavour observables in the context of the minimal model will be published
soon \cite{inprep}.

\section{The Minimal Model}
In the minimal realization of the RS scenario all SM fields except for the
Higgs are five-dimensional (5D) fields. In
order to solve the gauge hierachy problem, the Higgs must be confined to (or
localized
close to) the IR brane, where the UV cutoff becomes of
$\ord\left(\text{few TeV}\right)$,
due to the warp factor $\epsilon\equiv e^{-kr\pi}\approx 10^{-16}$.

Introducing a coordinate $t=\epsilon\,e^{\sigma(\phi)}$ along the
extra dimension \cite{Grossman:1999ra}, which runs from $t=\epsilon$
on the UV brane to $t=1$ on the IR brane, the KK
decompositions of the left-handed (right-handed) components of the
five-dimensional $SU(2)_L$ doublet (singlet) quark fields read
\begin{eqnarray}\label{KKdecomp}
   q_L(x,t)
   &\propto \mbox{diag}\left[ F(c_{Q_i})\,t^{c_{Q_i}} \right]
    \boldsymbol{U}_q\,q_L^{(0)}(x) + \ord\bigg(
\displaystyle{\frac{v^2}{\Mkk^2}} \bigg)+ \mbox{KK
modes} \,, \notag\\
   q_R^c(x,t)
   &\propto \mbox{diag}\left[ F(c_{q_i})\,t^{c_{q_i}} \right]
    \boldsymbol{W}_q\,q_R^{(0)}(x) + \ord\bigg(
\displaystyle{\frac{v^2}{\Mkk^2}} \bigg) +
\mbox{KK modes} \,,
\end{eqnarray}
where $q=u,d$ stands for up- and down-type quarks, respectively. The fields
are three-component vectors in flavor space. 5D
fields on the left-hand side refer to interaction eigenstates, while
the four-dimensional (4D) fields appearing on the right-hand side are mass
eigenstate. The superscript ``(0)'' denotes the so-called ``zero modes'', which
correspond to the light SM
fermions. Heavy KK fermions will not play a role in the following.

The ``zero-mode'' profiles
$F(c_{Q_i,q_i})$ 
are exponentially suppressed in the volume factor $L\equiv -\ln \epsilon$ if
the
bulk mass
parameters { $c_{Q_i}=+M_{Q_i}/k$} and $c_{q_i}=-M_{q_i}/k$ are
smaller than the critical value $-1/2$, in which case $F(c)\sim
e^{L(c+\frac12)}$ \cite{Grossman:1999ra, Gherghetta:2000qt}. Here
$\boldmath{M}_Q$ and $\boldmath{M}_{u,d}$ are the mass matrices of the
5D $SU(2)_L$ doublet and singlet fermions. 
This
mechanism explains in a natural way the large hierarchies observed in
the spectrum of the quark masses \cite{Gherghetta:2000qt,
  Huber:2000ie}, which follow from the eigenvalues of the effective
Yukawa matrices
\begin{equation}
\label{Yeff}
   \boldsymbol{Y}_q^{\rm eff} = \mbox{diag}\left[ F(c_{Q_i}) \right]
   \boldsymbol{Y}_q\,\mbox{diag}\left[ F(c_{q_i}) \right]
   = \boldsymbol{U}_q\,\boldsymbol{\lambda}_q\,\boldsymbol{W}_q^\dagger
\,,\quad\\[3pt]
\end{equation}and are up to $\ord(1)$ factors
\begin{equation}\label{qmasses}
m_{u_i}\sim\frac{v}{\sqrt{2}}|F(c_{Q_i})F(c_{u_i})|\,,\qquad
m_{d_i}\sim\frac{v}{\sqrt{2}}|F(c_{Q_i})F(c_{d_i})|\,.
\end{equation} 
The 5D Yukawa matrices $\boldsymbol{Y}_q$ are assumed to have
$\ord(1)$ complex entries, and $\boldsymbol{\lambda}_q$
are diagonal matrices with entries $(\lambda_q)_{ii}=\sqrt2
m_{q_i}/v$. The unitary matrices $\boldsymbol{U}_q$ and $\boldsymbol{W}_q$
appearing
in (\ref{KKdecomp}) and (\ref{Yeff}) have a hierarchical structure
given by quotients of the zero-mode profiles.

The profiles of the SM weak gauge bosons
receive $t$-dependent corrections due to electroweak symmetry
breaking, whereas
the massless gluon and photon modes remain flat
along the extra dimension. 
%Up to an overall normalization one finds
%\cite{Csaki:2002gy}
%\begin{equation}\label{eq:chi0}
%   \chi_0(t)\propto 1 - \frac{x_0^2\,t^2}{2} 
%   \left( L - \frac12 + \ln t \right) + \ord(x_0^4) \,,
%\end{equation}
%where $x_0=m_{W,Z}/\Mkk$. Overlap integrals of the $t$-dependent terms
%with fermion profiles give rise to flavor-changing effects.
 In order to compute
the full contribution to tree-level processes one has to consider the whole
tower of KK modes. 
The sum over the KK tower of gauge bosons can be evaluated 
by generalizing a method developed in \cite{Hirn:2007bb}. Dropping irrelevant
$\ord(\epsilon^2)$
constant terms, one finds for the sum over massive and massless KK gauge
bosons respectively
\begin{align}\label{KKsum}
 &\sum_{n}\,\frac{\chi_n(t)\,\chi_n(t')}{m_n^2}
   =\notag\\[10pt] &\begin{cases}
     \displaystyle{ \frac{1}{2\pi m_{W,Z}^2}} + \frac{1}{4\pi\Mkk^2} 
   \left[ L\,t_<^2 - L \left(t^2
+t'^2\right)+1-\frac{1}{2L}+\ord\left(\frac{m_{W,Z}^2}{\Mkk^2}
\right)
    \right] ,\\[20pt]
    \displaystyle{\frac{1}{4\pi\Mkk^2} }
   \left[ L\,t_<^2 - t^2 \left( \frac12 - \ln t \right)
    - t'^2 \left( \frac12 - \ln t' \right) + \frac{1}{2L} \right] ,
\end{cases}
\end{align}
where $t_<^2\equiv{\rm min}(t^2,t'^2)$. 
\\
In principle, the terms proportional to $t$ and $t'$ could cause dangerously
large FCNCs, but the corresponding vertices receive suppressions from the
zero-mode profiles of the associated fermions, mitigiating these effects. This
mechanism is known as the RS-GIM mechanism
\cite{Agashe:2004ay,Agashe:2004cp,Agashe:2005hk}.

\begin{figure}[t]
\hspace*{-\figureoverhang}%
\begin{minipage}{
\textwidth+\figureoverhang+\figureoverhang}%
\qquad\quad\includegraphics[bb= 0 -60 170 80]{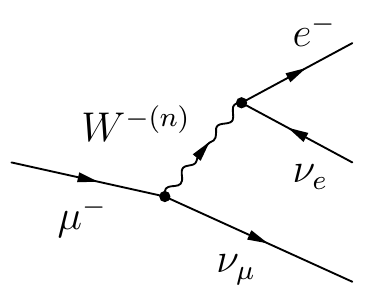}\hfill
\includegraphics[width=0.5\textwidth]{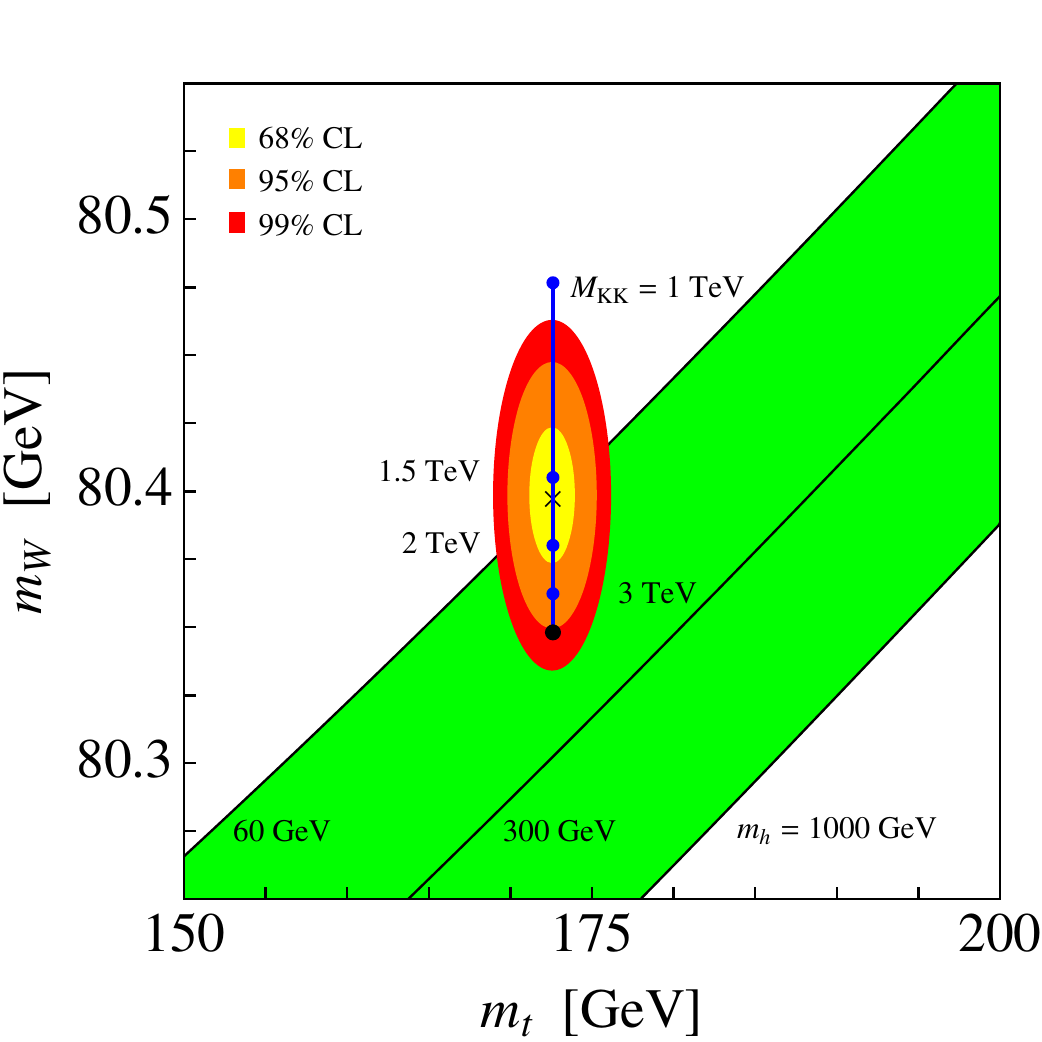}
\label{Wmass}
\caption{The left panel shows the tree-level diagram for $\mu^-\to
e^-\nu_{\mu}\bar{\nu}_{e}$, including the KK modes $W^{-(n)}$. The right panel
shows different probability regions from the direct measurement of
$m_W$ and $m_t$ at LEP2 and Tevatron, the SM prediction based on the value
of $G_F$ as a black dot and the SM expectation for
$m_h\in\left[60,1000\right]$ GeV as a green band. The RS prediction approaches
the SM for increasing $\Mkk$ as one follows the blue line.
}
\end{minipage}
\end{figure}

\section{Modification to the $\boldsymbol{W}$-Boson Mass}
In the SM the value from the direct measurement of the $W$-boson mass,
following from the
latest results of LEP2 and the Tevatron \cite{LEPEWWG:2005ema, Yao:2006px},
differs from the indirect
extraction from precise measurements of $\alpha, G_F,$ and $\sin^2\theta_W$ by
roughly $50 $ MeV. In the RS model, $G_F$, extracted from muon
decay, receives a universal contribution\footnote{In general, the $t$-dependent
part of \eqref{KKsum} also contributes, but they are strongly suppressed due to
the UV localization of the leptons. } from the exchange of the KK excitations
of the $W$ boson. The process is illustrated in the left panel of Figure
\ref{Wmass}. With \eqref{KKsum} one finds
\begin{equation}
 \frac{G_F}{\sqrt{2}}=\frac{g^2}{8m_W^2}\left[1+\frac{m_W^2}{2\Mkk^2}
\left(1-\frac{1}{2L}\right)+\ord{\left(\frac{m_{W,Z}^4}{\Mkk^4}\right)}\right]\,
.
\end{equation} 
This translates into a modification for the mass of the $W$ boson through the
SM relation
\begin{equation}
 \big(m_W^2\big)_{ind}\equiv \frac{\pi\alpha}{\sqrt2 G_F \sin^2\theta_W}\,,
\end{equation} 
so that 
\begin{equation}
 \big(m_W\big)_{ind}=m_W\left[1-\frac{m_W^2}{4\Mkk^2}\left(1-\frac{1}{2L}
 \right)+\ord\left(\frac{m_{W,Z}^4}{\Mkk^4}\right)\right]\,.
\end{equation} 
In the plot on the right panel of Figure \ref{Wmass} shows that the RS
prediction can therefore explain the difference for KK mass scales slightly
above
$1.5$~TeV, while allowing for a heavier Higgs mass, $m_h=400$ GeV ($m_h=1000$
GeV), the KK mass scale can even be lowered to $1.5$ TeV ($1$ TeV).

\begin{figure}
\hspace*{-\figureoverhang}%
\begin{minipage}{
\textwidth+\figureoverhang+\figureoverhang}
\includegraphics[width=0.45\textwidth]{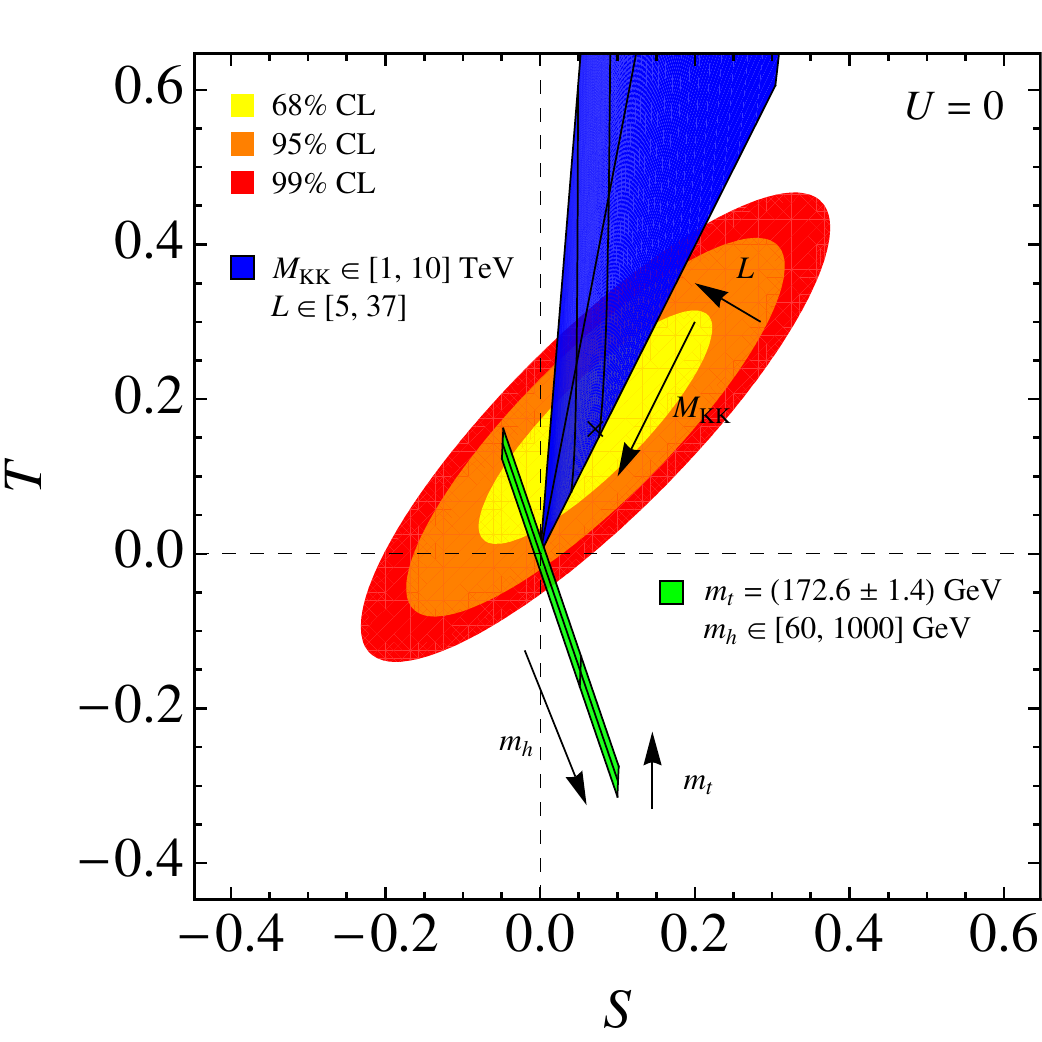}\hfill
\includegraphics[width=0.45\textwidth]{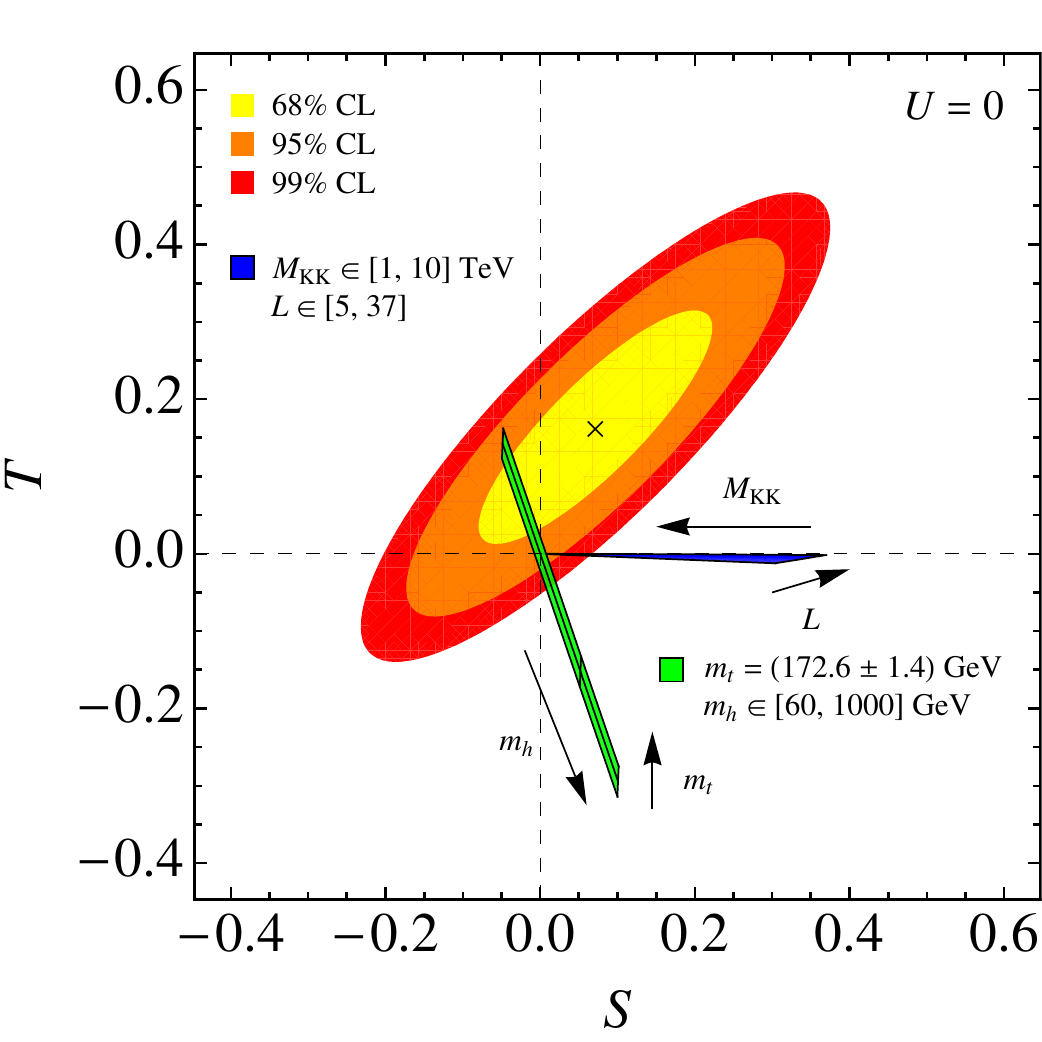}
\label{STU}
\caption{The left/right plot
shows different probability regions from a global fit to LEP and SLC
measurements for $S$ and $T$ in the RS model with/without custodial protection.
The
green stripes indicate the SM corrections for increasing $m_h\in[60,1000]$ GeV
and
$m_t=(172.6\pm1.4)$ GeV. The RS corrections are indicated by the blue shaded
area
and depend on the value of $L\in[5,37]$ and $\Mkk\in[1,10]$ TeV. }
\end{minipage}
\end{figure}

\section{$\boldsymbol{S}$, $\boldsymbol{T}$, and $\boldsymbol{U}$ Parameters}
Shifts from the SM values of the $S, T$ and $U$ parameters induced by new
physics indicate deviations from the electroweak radiative corrections expected
in the SM. In general, theories with additional heavy bosons call for an
extension of this setup \cite{Barbieri:2004qk}. But the additional parameters
include second
derivatives of vacuum polarization amplitudes and therefore turn out to be
very small. Measurable corrections are only found in $S$ and $T$
\cite{Carena:2003fx, Delgado:2007ne}
\begin{equation}
 S=\frac{2\pi v^2}{\Mkk^2}\left(1-\frac{1}{L}\right),\qquad T=\frac{\pi
v^2}{2\cos^2\theta_W \Mkk^2}\left(L-\frac{1}{2L}\right)\,.
\end{equation}  
As one can see from the left panel of Figure \ref{STU}, the correction to $T$
strongly constrains the parameter space of the minimal RS model and pushes
the KK mass scale up to $\Mkk> 4.0$ TeV. \\
There are three ways to solve this issue.\footnote{In fact there is at
least another one provided by large brane-localized kinetic terms, which is
however not discussed in this review.} As a first option,
one could assume a large Higgs mass. This corresponds to a negative shift
$\Delta T\sim \log m_h/m_h^{\mathrm{ref}}$ and can lower the bound, for
$m_h=1$~TeV, to $\Mkk>2.6 $~TeV.\footnote{Where the
reference value is set to $m_h=150$
GeV.} A relaxation can also be achieved by lowering the volume factor to
about $L=5$. That means abandoning the solution to the full hierachy
problem and requires a UV completion at
$\Lambda_{\mathrm{UV}}\approx 10^{3}$~TeV, but lowers the bound on the KK mass
scale to $\Mkk>1.5$~TeV. These models are called ``little RS'' scenario and were
first proposed in
\cite{Davoudiasl:2008hx}. 
A third possibility is to introduce an extended bulk symmetry group, a
so-called custodial symmetry $SU(3)_c\times SU(2)_L\times SU(2)_R\times
U(1)_X$. The tree-level corrections to $S$ and $T$ then read
\cite{Agashe:2003zs}
\begin{equation}
 S=\frac{2\pi v^2}{\Mkk^2}\left(1-\frac{1}{L}\right),\qquad T=-\frac{\pi
v^2}{4\cos^2\theta_W\Mkk^2}\frac{1}{L}\,,
\end{equation} 
and are displayed in the right panel of Figure \ref{STU}. While this can
provide a KK mass scale as low as $\Mkk=2.4$ TeV, it should be mentioned that
in this case a
large Higgs mass would spoil the electroweak fit. 

\section{$\boldsymbol{Z b\bar{b}}$ Couplings}
\begin{figure}
\hspace*{-\figureoverhang}%
\begin{minipage}{
\textwidth+\figureoverhang+\figureoverhang}
\includegraphics[width=0.45\textwidth]{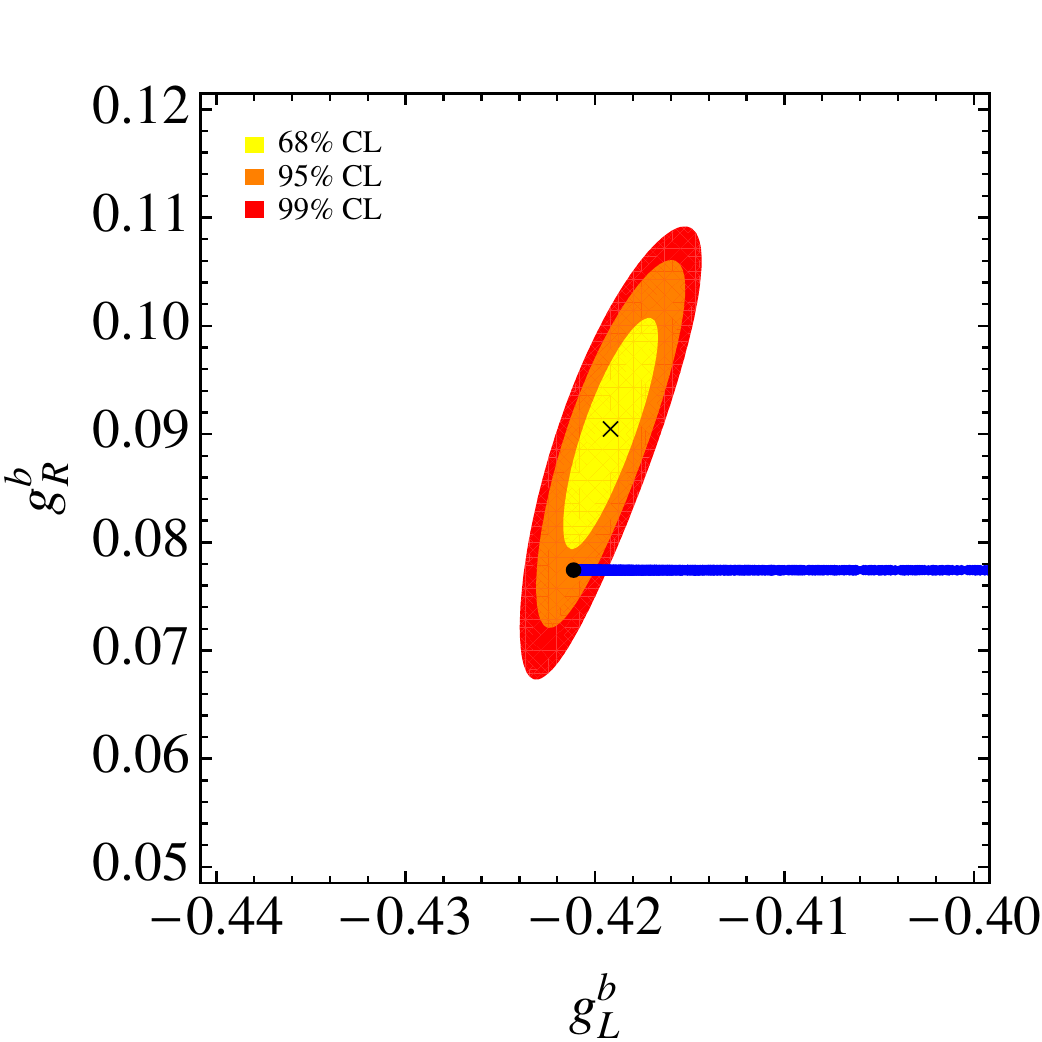}\hfill
\includegraphics[width=0.45\textwidth]{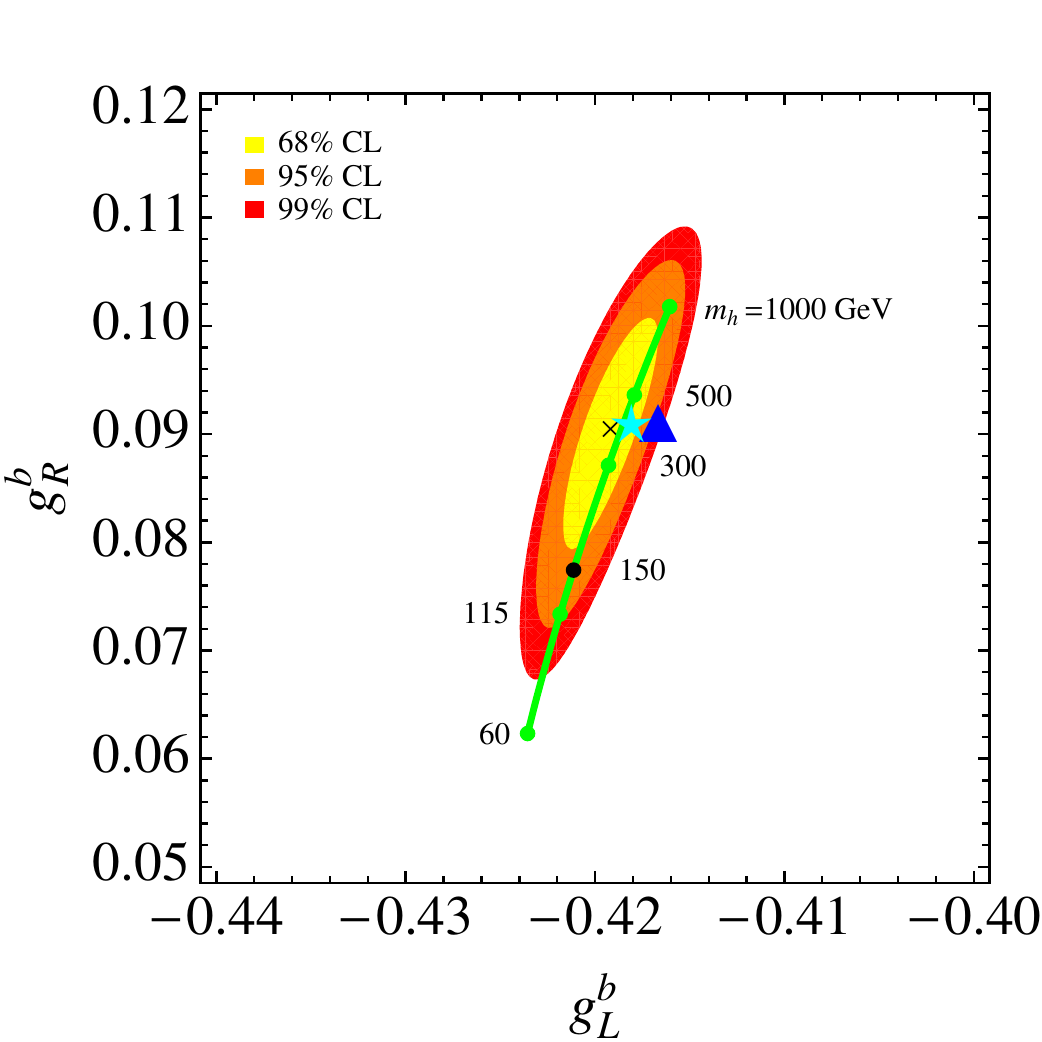}
\label{Zbb}
\caption{Both plots show probability regions for the experimentally extracted
value.$^{\ref{ftn}}$ The SM prediction is indicated by the black dot. RS points
lie on the blue horizontal stripe. The right panel shows that following the
green line, one can shift this stripe vertically by increasing $m_h\in[60,1000]$
GeV. The triangle and star indicate reference points at $\Mkk=1.5$ TeV and
$\Mkk=3$~TeV, respectively.}
\end{minipage}
\end{figure}

Another strong bound comes from the non-universal corrections to the coupling
of the $Z$ to bottom quarks. The corresponding
couplings read 
\begin{align}
  &\hspace{-9.5pt}g_L^b
   = \big(g_L^b\big)_{\mathrm{SM}}
    \left[ 1 - \frac{m_Z^2}{2\Mkk^2}\,
    \frac{F^2(c_{Q_3})}{3+2c_{Q_3}} \left(
    L - \frac{5+2c_{Q_3}}{2(3+2c_{Q_3})} \right)
\right]+\ord\left(\frac{m_b^2}{\Mkk^2}\right),\\[8pt]
   &\hspace{-9.5pt}g_R^b
   = \big(g_R^b\big)_{\mathrm{SM}}
    \left[ 1 - \frac{m_Z^2}{2\Mkk^2}\,
    \frac{F^2(c_{d_3})}{3+2c_{d_3}} \left(
    L - \frac{5+2c_{d_3}}{2(3+2c_{d_3})} \right)
\right]+\ord\left(\frac{m_b^2}{\Mkk^2}\right).
\end{align}
Unfortunately, as one can see in the left panel of Figure \ref{Zbb} the
right-handed coupling remains practically unaffected, while large corrections to
the left-handed coupling are possible. In order to reverse this feature one
can rescale the zero-mode profiles $F(c_{Q_3})$ and $F(c_{b_R})$. From the
quark mass relations \eqref{qmasses} follows that this redistribution requires a
large value for $c_{u_3}$. However, if $c_{u_3}$ becomes to large one has to
sacrifice the explanation of the quark mass hierarchy relying on order one bulk
mass parameters. While this problem does not appear in custodially protected
models since corrections to the left-handed $Z$ couplings basically vanish,
the issue can also be solved within the minimal model if one assumes a heavy
Higgs boson. The effect of a large Higgs mass is displayed on the right panel of
Figure \ref{Zbb}. Good agreement with the experimental value can be
achieved for $m_h = 400$ GeV.

\addtocounter{footnote}{1}\footnotetext{\label{ftn}For details see
\cite{Casagrande:2008hr}.}

\section{Concluding Remarks}
The $W$-boson mass difference between direct and indirect measurements can be
explained within the minimal RS model with reasonably low KK mass scale. The
tensions in $T$ and $Z\bar{b}b$ call for large
$\Mkk>4$ TeV, but can be resolved by introducing a heavy Higgs boson allowing
for a KK mass scale as low as $\Mkk>2.6$ TeV. Since the cutoff on the IR brane
is around the TeV scale, a Higgs mass of this
order is naturally expected in this model. A considerably lower Higgs mass would
introduce a little hierarchy problem. An alternative way to deal with these
issues is to introduce a custodial protection. However, a possible problem of
the latter model is that in the presence of a heavy Higgs boson a good
agreement with electroweak fits is challenging.

\section*{Acknowledgements}
I want to thank Sandro
Casagrande, Florian Goertz, Leonard Gr\"under, Uli Haisch, Matthias Neubert, and
Torsten Pfoh for useful and instructive discussions.

%\bibliographystyle{ijqc.bst}
%\bibliography{/home/mbauer/bibtexfiles/physicrefs.bib}

\end{document}